# Ultra-sensitive nanometric flat pigment for binocular stereoscopic image

Dejiao Hu[1#], Hao Li[1#], Yupeng Zhu[2], Yuqiu Lei[1], Jiajin Zheng[3], Yaoyu Cao[1], Bai-Ou Guan[1], Lei Bi[2*] and Xiangping Li[1*]

[1]Guangdong Provincial Key Laboratory of Optical Fiber Sensing and Communications, Institute of Photonics Technology, Jinan University, Guangzhou, China, 510632

[2]National Engineering Research Center of Electromagnetic Radiation Control Materials, University of Electronic Science and Technology of China, Chengdu, China, 610054

[3]College of Electronic and Optical Engineering & College of Microelectronics, Nanjing University of Posts and Telecommunications, Nanjing, China, 210046

* Email: xiangpingli@jnu.edu.cn; bilei@uestc.edu.cn

Two-dimensional (2D) transition metal dichalcogenides (TMDs) with tantalizing layer-dependent electronic and optical properties have emerged as a new paradigm for integrated flat opto-electronic devices. However, daunting challenges remain in deterministic fabrication of TMD layers with demanded shapes and thicknesses as well as light field manipulation in such atomic-thick layers with vanishingly small thicknesses compared to the wavelength. Here, we demonstrate ultra-sensitive light field manipulation in full visible ranges based on laser exfoliating $MoS_2$ layers with nanometric precisions. The nontrivial interfacial phase shifts stemming from the unique dispersion of $MoS_2$ layers integrated on the metallic substrate empower an ultra-sensitive resonance manipulation up to 12.8 nm per $MoS_2$ layer across the entire visible bands, which is more than five times larger than their counterparts. The interlayer van der Waals interactions endow a laser exfoliation method for on-demand patterning $MoS_2$ with atomic thickness precisions and subwavelength feature sizes in a facile and lithography-free fashion. With this, nanometric flat color prints and further binocular stereoscopic views by multi-perspective diffractive images can be realized. Our results with demonstrated practicality unlock full potentials and pave the way for widespread applications of emerging 2D flat optics.

**Introductions**

Since the first discovery[1,2], two-dimensional (2D) materials with exceptional optical and electronic properties have offered an unprecedented platform for developing ultra-compact flat opto-electronic devices with a considerable degree of miniaturization. Transition metal dichalcogenides (TMDs)[3,4] exhibiting sensitive layer-dependent properties including indirect-to-direct bandgap transition from bulk states to monolayers, emerge as a peculiar complementary to graphene for investigating excitonic light-matter interactions. As such, onrushing developments of extensive applications have been witnessed in photodetectors [5,6], valley emissions [7], transistors [8,9] and memories [10,11].

In spite of manifesting enticing high refractive indices[12,13], the light field manipulation in nanometric TMD layers remains dull sensitivity to their vanishingly small thickness compared to the wavelength of operation. Until very recently, nanostructured TMD layers at elevated thicknesses to hundreds of nanometres are starting to be appreciated as high-index dielectric resonators supporting distinct geometry-dependent Mie resonances, which starts a new research to shaping light's wavefronts by 2D materials based optical elements[14,15]. However, these demonstrations are achieved at the cost of degraded device compactness and integration. Moreover, the fabrication of nanostructured TMD layers mainly relies on mechanical or chemical exfoliation from bulk materials[16,17] and subsequently follows complex lithography procedures to produce demanded shapes and thicknesses[14,15]. Even though these approaches are demonstrated effective for fundamental researches, aforementioned challenges remain the major hurdle for the pragmatic and widespread applications of the emerging 2D flat optics.

Here, we demonstrate ultra-sensitive light field manipulation in resonance spectra by nanometrically maneuvering thickness of $MoS_2$ layers deposited on a gold substrate. By introducing the giant interfacial phase shifts associated with imaginary parts of their refractive indices, the resonance supported by $MoS_2$ layers can be tunable in the full visible range with an extreme sensitivity to the nanometric thickness. The interlayer van der Waals interactions enable laser induced exfoliation effects[18,19] which provides a pragmatic and lithography-free means to on-demand structure $MoS_2$ flakes at atomic layer precisions and subwavelength feature sizes. The proof-of-principle demonstration of flat color prints and further multi-perspective stereoscopic diffractive images unfolds the potential of 2D flat optics with practicality and up-scalability.

**Results**

**Ultra-sensitive resonance manipulation through laser exfoliating nanometric $MoS_2$ layers**. The typical configuration of nanometric $MoS_2$ multilayers integrated on a gold substrate for ultra-sensitive resonance manipulation by direct laser writing technique[20] is schematically illustrated in Fig. 1. The $MoS_2$ thin films with an initial thickness of 20 nm were deposited on a gold substrate (see Methods). A continuous wave (CW) laser beam at the wavelength of 532 nm was focused by an objective lens (50x, NA=0.75) to pattern $MoS_2$ layers. Light absorption in the upper layers can produce a local temperature rise that sublimates and burns out atoms in the vicinity of the focal region, which can be dexterously controlled by the laser recipe. Figure 1b showcases the reflection optical image of a tangram pattern fabricated through the laser writing method. The zones with distinct color appearances represent $MoS_2$ thin films with different nanometric thicknesses, verified by the topographic atomic force microscopy (AFM) image. Indeed, the structure exhibits extreme sensitivity in resonance tunability across in the full visible range through subtle variations of the thickness of $MoS_2$ thin films by merely tens of atomic layers.

To gain insights into the ultra-sensitive resonance manipulation, we first review the Fabry-Perot (FP) resonance supported by a thin film. FP resonance modes are standing waves formed by light wave propagating back and forth between two reflective interfaces, where constructive interference occurs after the light wave travels a round-trip. The resonance condition can be matched once the total phase accumulation including both propagation phase shifts and interfacial phase retardations acquired at interfaces to be an integer modulus of $2\pi$. Thus, the resonance wavelength can be simply derived as

$$\lambda = \frac{2n}{m - (\frac{\varphi_1 + \varphi_2}{2\pi})} h, \qquad (1)$$

where *n* is the real part of the refractive index of the FP layer, h is its thickness, m is the order of resonance, and $\varphi_1$ and $\varphi_2$ are interfacial phase shifts at the two interfaces, respectively. It can be seen that the sensitivity of the resonance wavelength to the FP thickness is mainly governed by the refractive index of the material and total interfacial phase shifts at the two interfaces. Especially, when the total interfacial phase shifts are reaching 2π, the denominator will approach zero for the first order resonance (*m*=1). In this case, the resonance wavelength will be ultra-sensitive to even nanometric thicknesses of the FP medium, which leads to performances superior to conventional optical coatings relying on the propagation phase accumulation in a quarter-wave-thick film [21] (Supplementary Note and Fig. S1).

To realize such ultra-sensitive resonance manipulation, dielectric thin films with large complex indices as well as proper substrate designs are of vital importance. The unique dispersion of $MoS_2$ thin films integrated on the Au substrate empowers broadband interfacial phase shifts reaching 2π in the visible regime (Fig. 1c). Consequently, the largest tunability in reflectance spectra corresponding to a wide color palette by varying nanometric thickness of $MoS_2$ layers can be realized, which exceedingly outperforms the other substrates such as Al and Silicon (Supplementary Fig. S2). Figure 1d depicts the theoretical calculation results of the reflectance spectra of the $MoS_2$-Au structure with different numbers of layers (see Methods). As can already be inferred from the white dotted line marking out the evolution of resonance wavelengths, nearly 400 nm shift from 800 nm to 430 nm covering the entire visible light can be obtained by a thickness variation from 30 layers to a monolayer. It is worth noting that even the extinction-associated interfacial phase shifts[22] are ubiquitous among dissipative materials with a large imaginary part of refractive index, the enticing dispersion properties of $MoS_2$ thin films covering the outermost range in the complex index diagram (Fig. 1c) manifest a far superior sensitivity of 12.8 nm per $MoS_2$ layer, which showcases more than 5 times and 2 times greater than that of 2.3 nm per graphene layer and 5.9 nm per Ge layer, respectively (Supplementary Fig. S3).

Apparently, the interlayer interaction by van der Waals forces results in the dominant heat dissipation along the in-plane direction rather than the out-of-plane direction[18,23]. The local temperature in the upper layers is quickly built up to exceed the burning temperature and the beneath layers are less susceptible unless irradiation at high powers. This provides the basis for reliable thickness control at atomic layer precisions by establishing a standardized laser exfoliating recipe. Figures 2a to d illustrate the systematic investigation of nanometric thickness control at variant laser doses, and characterization obtained through AFM and Raman spectroscopy (see Methods). The laser scanning speed is optimized and fixed at 0.1 mm/s in the whole experiment (see Methods). By varying the laser irradiance powers with an increment of 8 mW, the thickness of $MoS_2$ thin films can be reduced in a staircase behavior with a step height of approximately 3 nm corresponding to 4 layers (given the monolayer thickness about 0.67 nm[24]). The surface roughness is on the order of 7.491 nm due to the presence of splashed or unremoved residuals, albeit no significant influence on their light field manipulation capabilities.

Since the frequency interval between $E_{2g}$ and $A_{1g}$ Raman modes depends monotonically on the number of $MoS_2$ layers[25,26], Raman spectroscopy is utilized to characterize the laser exfoliation process. Figure 2c depicts the evolution of Raman spectroscopy during the laser exfoliation. Initially, the frequency interval is 26 cm$^{-1}$ for the 20 nm thick flakes indicating a multilayer state. As the laser dose increases, the frequency interval monotonically decreases close to 21 cm$^{-1}$ implying a monolayer to bilayer state. The above characterization consolidates the effectiveness of the laser writing technique for thickness control at atomic precisions as well as up-scalability for on-demand fabricating layered MoS$_2$ based optical elements.

From Eqs. (1), it clearly reveals that the nanometric thickness variation of MoS$_2$ thin films integrated on the Au substrate leads to an ultra-sensitive light field management in resonance wavelengths. Figures 2e and f showcase the theoretical calculation results of reflection spectra from that structure at variant nanometric thicknesses of MoS$_2$ thin films as well as corresponding experimental results. The experiments are in good congruence with the theoretical predictions. It can be seen that the reflection valley caused by the resonant absorption can be continuously tuned in the whole range of visible light from the wavelength of more than 700 nm to 500 nm, by merely a thickness decrease from approximately 20 nm to 5 nm. Optical micrographs of the colors generated from the laser thinned zones and corresponding color coordinate map are shown in Figs. 2g and h, respectively.

**Flat pigments and binocular stereoscopic images**. The ultra-sensitive resonance control allows the generation of nanometric flat pigments for color images[27-30](Supplementary Fig. S4). Figures 3a to e are collections of halftone images patterned by laser writing methods. The color halftone effect is realized through controlling the printed pixel density. The underlying mechanism of laser sublimation can be attributed to photothermal effects, where the local heating effect and heat diffusion can be flexibly controlled by the focusing conditions to pattern subwavelength feature sizes.

Through dexterous control of the laser recipe, the patterned image composed of subwavelength scale pixels can achieve a high spatial resolution up to 58,000 dpi. The minimum pixel size is measured around 400 nm given a focusing lens with a numerical aperture of 0.75. To demonstrate the up-scalability, a *Canton Tower* with a millimeter scale was printed with a high resolution and high fidelity (Fig. 3a). Scanning electron microscope (SEM) image of the prints is depicted in Fig. 3b. The SEM image and optical micrograph of the selected region with fine details are shown in Figs. 3c and d, respectively. Figs. 3f to g show another example of images or even color QR codes printed with continuous tone. More examples of color images generated by nanometric flat pigments can be seen in Fig. S5.

Not only for sensitive resonance wavelength modulation to generate flat pigments, this nanometric thickness dependent resonance control can also introduce efficient diffraction effects for shaping the impinging wavefronts. Figure S6 illustrates the measured diffraction efficiency of patterned micro-gratings with a period of 1000 nm and a thickness contrast of 10 nm as a function of wavelengths. Given the ultra-thin nanometric thickness of the top $MoS_2$ thin films, the first-order diffraction efficiency can reach up to 36% at the optimal wavelength of 560 nm, which is sufficient to develop a new binocular stereoscopic view approach by multi-perspective diffractive images. The principle and the configuration are schemed in Figs. 4a and b. Two perspective images projected at two different directions were encoded into two sets of diffractive gratings with different periods of 1200 nm and 1800 nm, respectively (Supplementary Fig. S7). By horizontally interleaving the two sets of laser printed gratings to form a 10 µm sized supercell, two perspective images can be diffracted and projected to different directions corresponding to left and right eyes at an oblique incidence. Figures 4c and d showcase the captured portrait images taken from different perspective that can form a binocular stereoscopic view. The zoom-in views of the SEM image and optical micrograph of the laser patterned interleaved grating structures are shown in Figs. 4e and f.

**Discussions**

By removing complex and sophisticated electron beam nanolithography procedures that are heavily resorted to, the demonstrated laser writing methods with atomic thickness precisions and subwavelength feature sizes paves the way to on-demand fabrication of ultra-thin TMD based opto-electronic devices with great flexibility and up-scalability. The exotic dispersion of TMD materials empowers an ultra-sensitive light field manipulation scheme and opens avenues to a new class of nanometric flat optical elements. Furthermore, the proof-of-principle demonstration of nanometric flat pigments and further multi-perspective stereoscopic views provides a viable scheme to construct light field manipulation based flat optics with ultra-compact footprints and superior miniaturization. Combing their layer-dependent behaviors of bandgaps, the ultra-sensitive light field management in resonance control by the nanometric thickness of TMD thin films can dramatically push the state-of-the-art and nourish unprecedented new functionalities of emerging 2D integrated opto-electronic devices.

**Methods**

**Sample preparation**. The MoS$_2$ thin film with a lateral size of several millimeters was firstly deposited onto Si/SiO$_2$ (300 nm SiO$_2$) substrate with pulse laser deposition (PLD) method, and then was transferred onto a gold substrate obtained by heat evaporation. In the PLD process, a KrF laser with 248 nm wavelength and the commercial MoS$_2$ target were used. The condition of fabrication MoS$_2$ layer was 10$^{-5}$ pressure and 780 degree. After the deposition, the pressure was maintained and annealed for 5 minutes at the same condition, and then the temperature was decreased at a rate of 10° C/min to the room temperature. The power of the laser pulse was 150 mJ with a repeat frequency of 5 Hz. The thickness of the MoS$_2$ layer was controlled by the pulse number of the laser. The MoS$_2$ film was transferred onto the gold substrate with wet transfer method, where a polymethylmethacrylate (PMMA) layer was spinned onto the Si/SiO2/MoS$_2$ films with a spin rate of 500 rmp for 10 s and then 3000 rpm for 60 s. And then, the PMMA layer was immobilized through baking for 10 mins at a temperature of 100° C. After this, the sample was immersed in KOH solution to remove the substrate and the MoS$_2$ with PMMA film float onto the solution surface. Then the residual KOH was washed by putting the MoS$_2$ and PMMA film in deionized water for several seconds. The MoS$_2$ and PMMA film was then spread onto the gold substrate and baked for 10 mins under a temperature of 95° C. Then the PMMA layer was removed by immersing in acetone, ethanol and deionized water for 30 s respectively.

**Laser printing**. The samples were placed on a computer-controlled 3D translation stage. The continuous wave laser beam at the wavelength of 532 nm was attenuated and focused by an objective lens (×50, 0.75NA). The beam power was adjusted to the desired value with a neutral density attenuator. During the laser printing, the movement of the translation stage was synchronized with the laser shutter to control the exposure of irradiances.

**Sample characterization and optical measurements**. Reflection color images of the sample were characterized using an objective lens (MPlanFL N, 50×/0.8, Olympus Co.). A CCD camera (Olympus, BX53, Olympus Co) was used to acquire the images from the sample. The spectra were characterized with a home built confocal microscope coupled to a spectrometer (Andori500). The sample was illuminated using a halogen white light source using an objective lens (MPlanFL N, 20×/0.45, Olympus Co.). The reflected light was collected through the same objective lens and recorded using a spectrometer. The reflected intensity was normalized by the spectrum of the lamp obtained by reflection measurements with a silver mirror.

**Characterization**. An atomic force microscope (Ntegra solaris, NT-MDT Spectrum Instruments, Moscow, Russia) has been used to study the topography and determine the height of patterned MoS$_2$ flakes. A Raman spectrometer (RENISHAW inVia) was used in a backscattering configuration excited with a visible laser beam (λ = 532 nm, power 5 mw) to confirm the layer number of MoS$_2$ flakes.

**Numerical calculations**. The reflection and transmission coefficients from a single interface is calculated by using the Fresnel equations, $r_{ij} = (p_i - p_j)/(p_i + p_j)$, $t_{ij} = 2p_i/(p_i + p_j)$, $p_i = n_i \cos(\theta_i)$. Here, $t_{ij}$ and $r_{ij}$ are transmission and reflection coefficients under illumination from medium i to medium j, $n_i$ is the complex refractive index of medium i and $\theta_i$ is the angle between the propagating direction of the light wave within the medium and the normal direction of the layered films. The reflection phase shift from the interfaces was extracted from the reflection coefficients and then were added up to the total phase shift. The reflection from the air-FP-substrate layers was directly calculated by using $r = r_{12} + \dfrac{t_{12} t_{21} r_{23} e^{-i2\varphi}}{1 - r_{21} r_{23} e^{-i2\varphi}}$, where 1, 2, 3 represent the air, FP layer and substrate, respectively, $\varphi$ is the propagation phase accumulation in the FP layer.

**Acknowledgements**

This research was supported by National Key R&D Program of China (YS2018YFB110012), Ministry of Science and Technology of the People's Republic of China (MOST) (Grant Nos. 2016YFA0300802, 2018YFE0109200), National Natural Science Foundation of China (NSFC) (Grant 61705084, 51972044), Guangdong Provincial Innovation and Entrepreneurship Project (Grant 2016ZT06D081), and Sichuan Provincial Science and Technology Department (Grant No. 2019YFH0154).

**Author contributions:** X. L. and B. L. conceived the idea and supervised the project. D. H., H. Li. and Y. L. performed the experiments. Y. Z. prepared the sample. D. H. and H. Li performed the modelling and simulation. D. H., H. Li., B. L. and X. L. analyzed data and prepared the manuscript. All authors contributed to the discussion and manuscript writing.

**Additional information:** Authors declare no competing interests. Supplementary information accompanies this paper on www.nature.com/naturematerials. Reprints and permissions information is available online at http://www.nature.com/reprints. Correspondence and requests for materials should be addressed to X.L. or B. L.

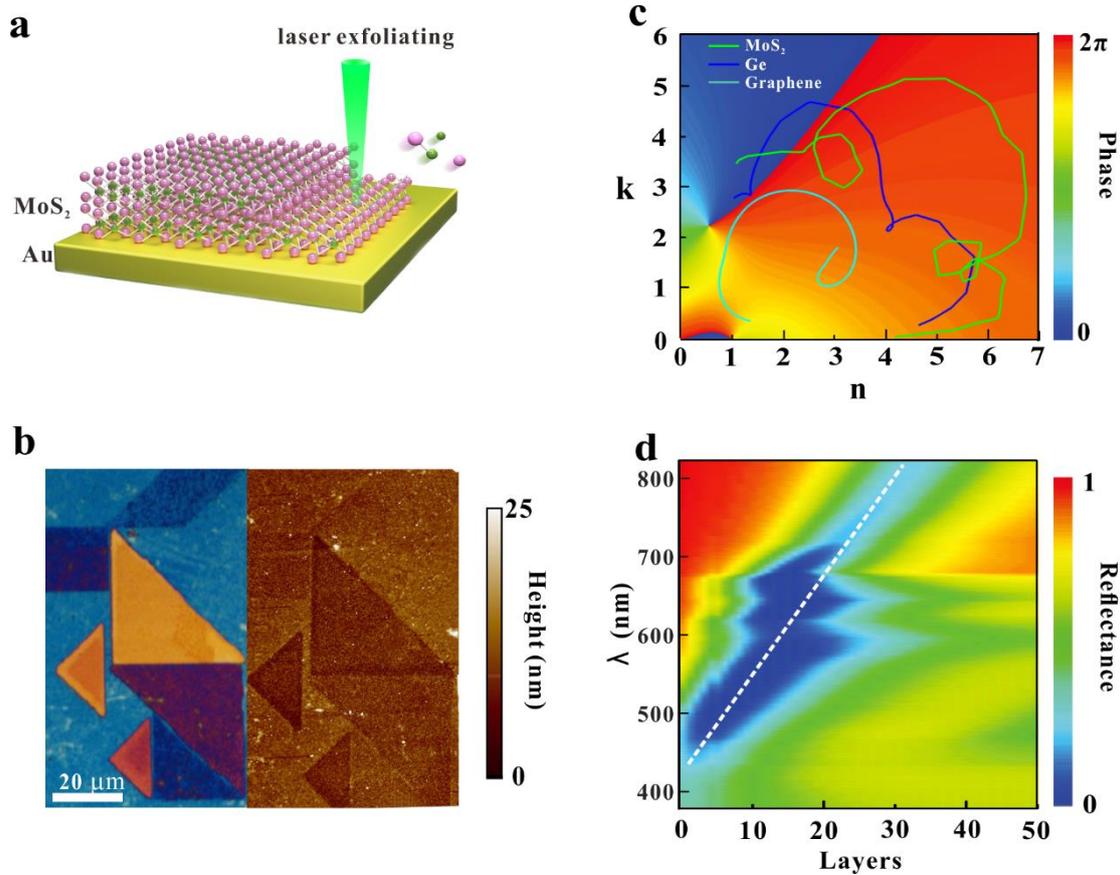

**Figure 1. Ultra-sensitive resonance manipulation by laser exfoliating MoS₂ thin films.**

**a**, Schematic diagram of tightly focused laser for exfoliating multilayer MoS$_2$ integrated on the Au substrate with atomic thickness precisions. **b**, Optical and the AFM images of a tangram pattern printed by the facile laser writing method. **c**, The calculated total interfacial phase shifts for a dielectric layer-Au configuration with variant complex refractive index overlaid with the complex refractive index diagram of MoS$_2$ (green), graphene (cyan), and Ge (blue). **d**, The calculated reflectance spectra of the MoS$_2$ thin films on the Au substrate with different numbers of layers.

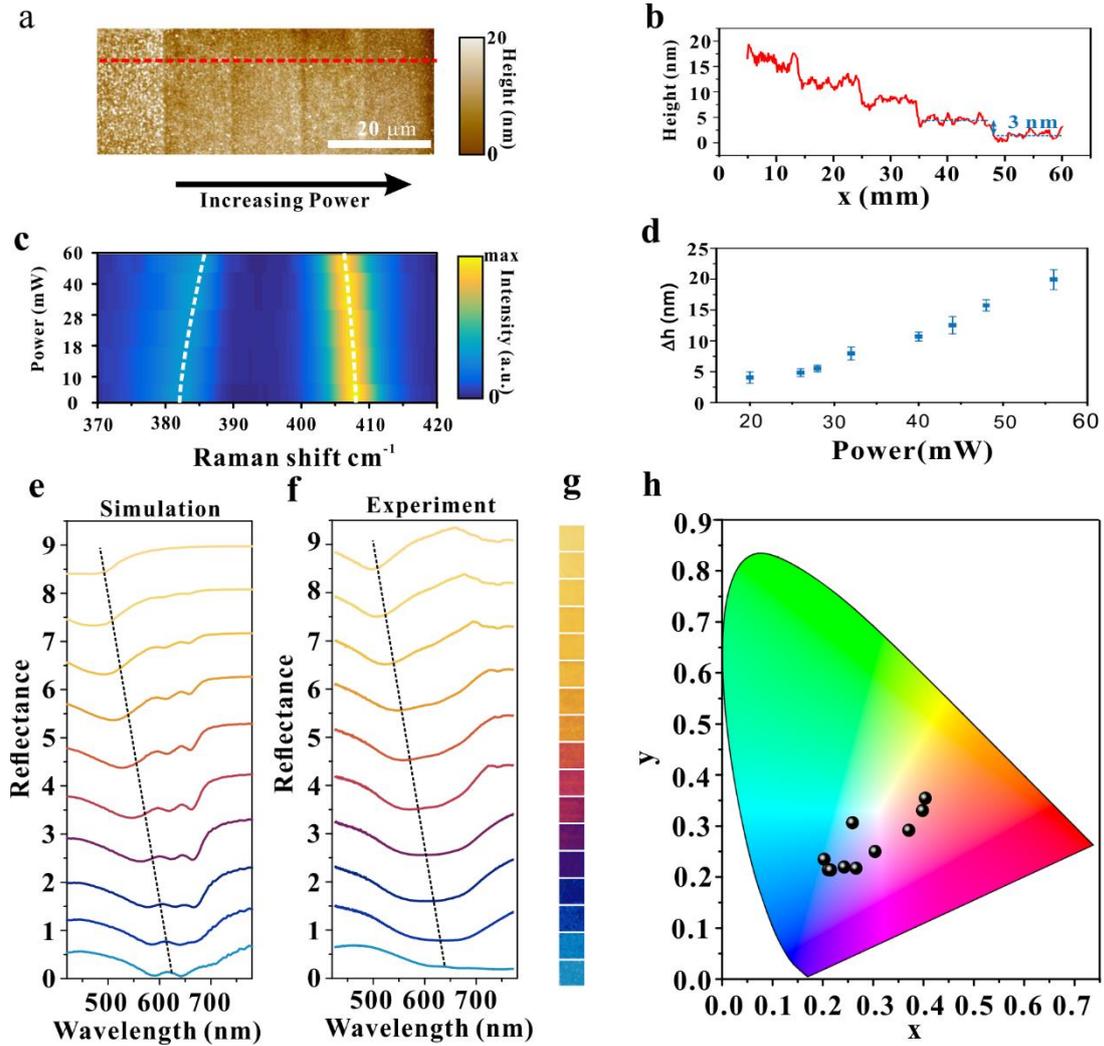

**Figure 2. Laser exfoliating MoS₂ thin films with nanometric thickness precisions. a**, The AFM image of laser thinned regions obtained at different laser doses. **b**, The height change corresponding to the red dotted line in **a**. The height of the steps is about 3 nm. **c**, The variation of characteristic Raman spectra of MoS₂ flakes as the increase of laser powers. **d**, Extracted thickness change from Raman spectra at corresponding laser powers. **e** and **f**, Simulation and experimental results of reflectance spectra of MoS₂ thin films prepared on the Au substrate with different nanometric thicknesses. **g** and **h**, Experimentally obtained reflection color palettes and color coordinate diagram through laser exfoliating at variant powers.

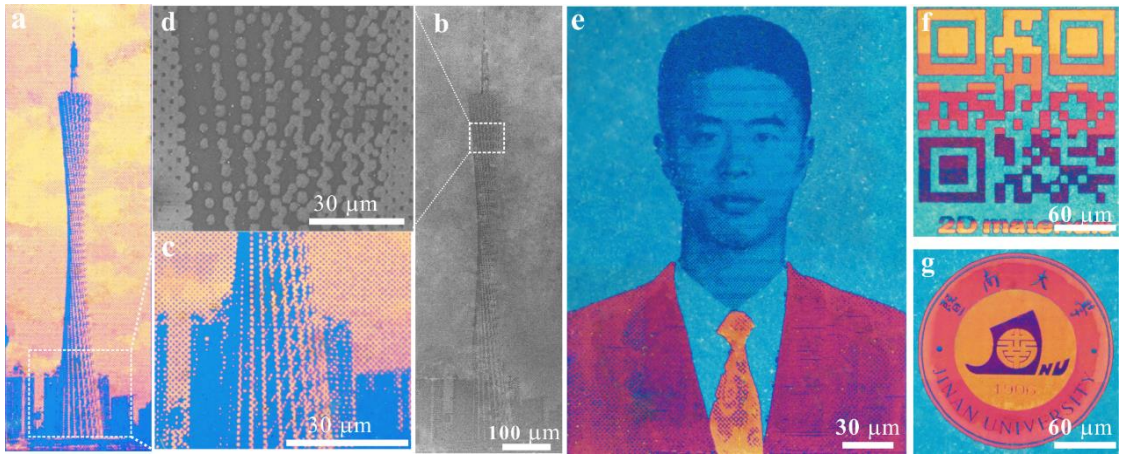

**Figure 3. High fidelity color images by laser exfoliated nanometric flat pigments. a** and **c** are optical microscopic image and zoom-in view. **b** and **d** are SEM image and sectional zoom-in view. A collection of color images printed by the laser exfoliation method with atomic thickness precisions and subwavelength feature sizes in continuous tone (**f, g**) and half tone modes (**a, e**) demonstrates the capability of high color fidelity and high spatial resolution.

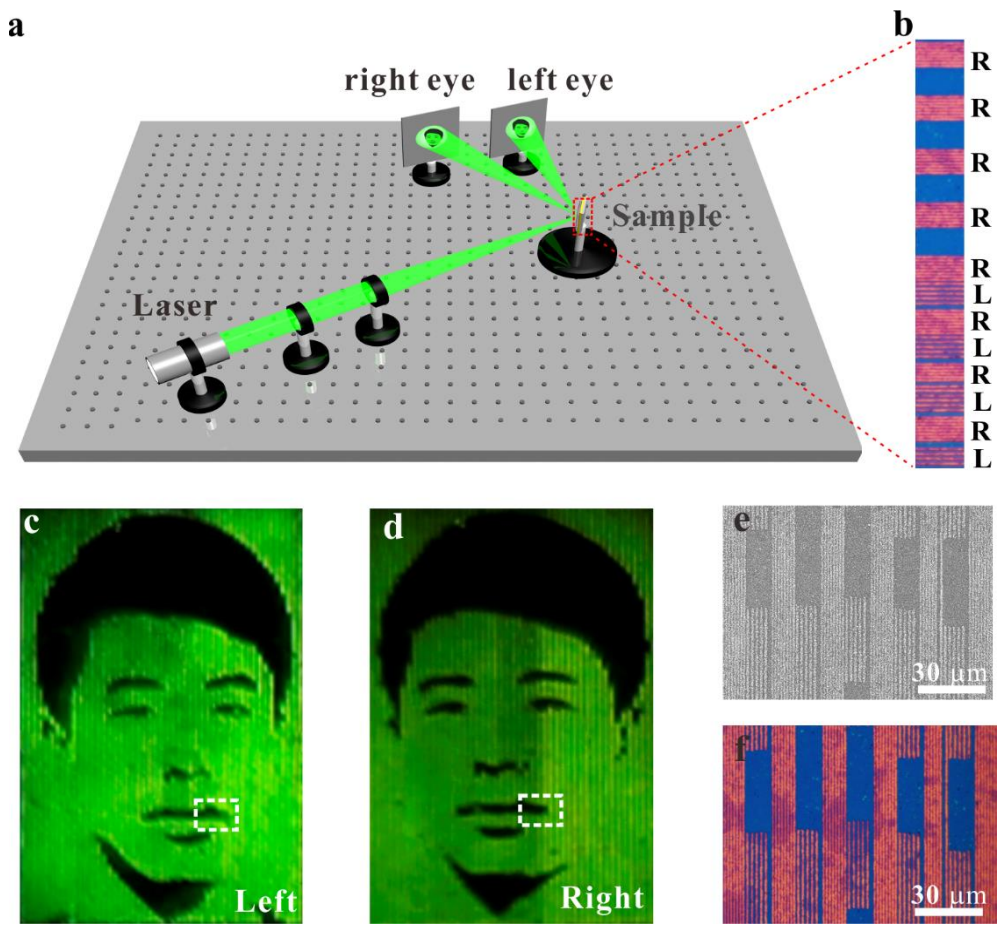

**Figure 4**. **Nanometric flat pigments for binocular stereoscopic views by multi-perspective diffractive images. a**, The optical configuration for binocular stereoscopic images. **b**, Principle of multi-perspective stereoscopic images through interleaved two sets of diffractive gratings. **c** and **d** are different images acquired by the left and right eyes, which were false-color images of the human face taken from different angles, respectively. **e** and **f**, Enlarged views of optical microscopic images and corresponding SEM image in the white dotted boxes in **c** and **d**.